\documentclass[]{aastex}

\usepackage{emulateapj5}
\usepackage{onecolfloat}
\usepackage{graphicx} 
\usepackage{fancyheadings} 
\usepackage{ulem}
\usepackage{rotating}
\usepackage{lscape}

\newcommand{\sqdeg}{\square^\circ}
\newcommand{\mnhi}{N_{\rm HI}}
\newcommand{\nhi}{$N_{\rm HI}$}
\newcommand{\ew}{W_\lambda}

\newcommand{\msol}{M_\odot}

\newcommand{\delv}{\Delta v}

\newcommand{\lya}{Ly$\alpha$}

\newcommand{\kms}{km~s$^{-1}$ }
\newcommand{\cm}[1]{\, {\rm cm^{#1}}}
\newcommand{\N}[1]{{N({\rm #1})}}

\newcommand{\sci}[1]{{\rm \; \times \; 10^{#1}}}

\newcommand{\perd}{\;\;\; .}

\newcommand{\mkms}{{\rm \; km\;s^{-1}}}

\begin{document}

\twocolumn[%
\submitted{Accepted to ApJ: January 30, 2006}

\title{Probing the IGM/Galaxy Connection Toward PKS0405-123 III: 
The Galaxy Survey and Correlations with \ion{O}{6} Absorbers}

\author{Jason X. Prochaska\altaffilmark{1},
	Benjamin J. Weiner\altaffilmark{2},
	Hsiao-Wen Chen\altaffilmark{3},
	and John S. Mulchaey\altaffilmark{4}}

\begin{abstract} 

We present a galaxy survey of the field surrounding PKS0405--123
performed with the WFCCD spectrometer at Las Campanas Observatory.
The survey is comprised of two datasets: (1) a greater than
95\% complete survey to $R = 20$\,mag of the field centered
on PKS0405--123 with $10'$ radius ($L \approx 0.1 L_*$ and radius
of 1~Mpc at $z=0.1$); 
and (2) a set of four discontiguous (i.e.\ non-overlapping),  
flanking fields covering $\approx 1\square^\circ$ area
with completeness $\approx 90\%$ to $R=19.5$\,mag.
With these datasets, one can examine the local and large-scale
galactic environment of the absorption systems identified toward
PKS0405--123.  
In this paper, we focus on the \ion{O}{6} systems analyzed in Paper~I.
The results suggest that this gas arises in a diverse set of
galactic environments including the halos of individual galaxies,
galaxy groups, filamentary-like structures,
and also regions devoid of luminous galaxies.
In this small sample, there are no obvious trends between galactic
environment and the physical properties of the gas.  Furthermore,
we find similar results for a set of absorption systems with
comparable \nhi\ but no detectable metal-lines.  The observations
indicate that metals are distributed throughout a wide range
of environments in the local universe.  Future papers in this
series will address the distribution of galactic environments
associated with metal-line systems and the \lya\ forest based on
data for over 10 additional fields.
All of the data presented in this paper is made public at a 
dedicated web site.


\end{abstract}
]

\altaffiltext{1}{UCO/Lick Observatory; University of California, Santa Cruz;
	Santa Cruz, CA 95064; xavier@ucolick.org}
\altaffiltext{2}{Department of Astronomy; University of Maryland; 
	College Park, MD 20742-2421; bjw@astro.umd.edu}
\altaffiltext{3}{Department of Astronomy; University of Chicago;
	5640 S. Ellis Ave., Chicago, IL 60637; hchen@oddjob.uchicago.edu}
\altaffiltext{4}{Observatories of the Carnegie Institution of
Washington, 213 Santa Barbara St., Pasadena, CA 91101; mulchaey@ociw.edu}

\pagestyle{fancyplain}
\lhead[\fancyplain{}{\thepage}]{\fancyplain{}{PROCHASKA ET AL.}}
\rhead[\fancyplain{}{Probing the IGM/Galaxy Connection Toward PKS0405-123 III:}]{\fancyplain{}{\thepage}}
\setlength{\headrulewidth=0pt}
\cfoot{}

\section{Introduction}

Since the discovery of quasar absorption line (QAL) systems, 
establishing and defining
their relationship to known galaxies has been a central
area of research.  Unfortunately, this pursuit has been hampered
by a number of factors.  The majority of QAL systems are identified
in optical quasar spectra at $z>2$ where galaxies
have very faint apparent magnitudes.  Furthermore, the glare of 
the background quasar prohibits imaging and spectroscopy at small
angular radii.  At $z \gtrsim 1$, authors have had modest
success by studying
\ion{Mg}{2} absorbers, metal-line systems selected by the doublet
feature at rest-frame wavelengths $\lambda=2796$ and 2803 \AA\ that 
are observable to redshift as low as $z=0.2$ in optical spectroscopic 
surveys.
The few studies to date associate the strongest systems with
the halos of $\approx L_*$ galaxies at $z\sim 1$ 
\citep{bergeron88,lzt93,steidel93}.
Even at these redshifts, however, it is difficult to conduct large
surveys to sub-$L_*$ luminosity to trace the galactic environment
of this specific class of QAL systems.

With the advance of UV spectroscopy on space-borne facilities, such as
the {\it Hubble Space Telescope}, QAL studies have
been carried out on a growing sample of low redshift AGN and quasars.
To date, the majority of studies
at $z<1$ have focused on the nature of strong \lya-forest
absorbers of neutral hydrogen column density $\log\,\mnhi \ge 14$ 
\citep[e.g.][]{morris93,lzt96,jannuzi98}
%(e.g.\ Morris \etal\ 1993; Lanzetta \etal\ 1998; Jannuzi \etal\ 1998), 
and some authors have also pushed the study to weak absorbers of 
$\log \mnhi =12-13.5$
along several lines of sight 
\citep{impey99,penton02,penton04}.
%(Impey et al.\ 1999; Penton et al.\ 2002, 2004).
Together these results show that on large-scales ($\sim 1$ Mpc) strong \lya\ 
absorbers are not distributed
randomly with respect to known galaxies, while weak absorbers are, 
exhibiting little/no
trace of large-scale galactic structure 
\citep[see also][]{grogin98,tripp98,bowen02,chen05}.
%(see also Grogin \& Geller 1998, Bowen \etal\ 2002; Chen \etal\ 2005) .

In addition, comprehensive 
surveys of galaxies and metal-line absorbers along common lines of sight 
have examined the correlation between galaxies and \ion{C}{4} systems 
with rest equivalent width $\ew \gtrsim 0.3$\AA\ \citep{chen01}.  
\cite{chen01} showed that nearly all galaxies found at impact parameter 
$\rho \leq 100 h^{-1}$\,kpc show associated \ion{C}{4} absorption while 
those at larger $\rho$ rarely do.  The authors interpreted 
this signature as a metal-enriched gaseous envelope surrounding galaxies,
i.e.\ the absorption is galactic as opposed to intergalactic. 
This interpretation was further supported by an apparent correlation
between the galaxy luminosity and \ion{C}{4} equivalent width.
It is worth noting that
\cite{adelberger03} have reported a similar correlation between
\ion{C}{4} absorption and luminous Lyman break galaxies at $z\sim 3$.
Furthermore, they describe a `halo' with physical radius $\rho \approx 
125$\,kpc (perhaps the result of starburst winds) in good agreement with
the \cite{chen01} result.  It remains an open question as to
whether this pair of results are a coincidence, a generic
characteristic of luminous galaxies, or even an 
indication of passive evolution 
in galactic halos from $z = 3$ to low redshift.

More recently, high resolution 
spectroscopy carried out at mid- and far-ultraviolet
wavelengths using the {\it Far Ultraviolet Spectroscopic Explorer} 
(FUSE) 
and the Space Telescope Imaging Spectrograph (STIS) on board HST has offered 
a unique avenue to study highly ionized gas at low redshifts based on 
observations of \ion{O}{6} $\lambda\lambda$1031,1037 
\citep[e.g.][]{savage98,tripp00,savage02}.
Several authors have examined the galactic environment
of \ion{O}{6} absorbers along the sightlines to PG1211$+$143 at 
$z_{\rm QSO}=0.081$ and PG1116$+$215 at $z_{\rm QSO}=0.177$ 
\citep{tumlinson05,sembach04}.  The results show that \ion{O}{6} absorbers
are correlated with large-scale galaxy distributions, but there is no
clear evidence to support an immediate connection to individual galactic
halos.

In this paper, we present optical spectra for $\sim 500$ objects 
in the field surrounding the quasar PKS0405$-$123.  In Paper~I 
\citep{pro04},
we analyzed the UV spectroscopy of this sightline obtained with 
FUSE and STIS.  We focused primarily on the metal-line systems
toward PKS0405$-$123, their ionization state and physical
properties.  
In Paper~II \citep{chen05}, we focused on the cross-correlation between 
galaxies and the \lya\ forest along the sightline 
\citep[see also][]{williger05}.
In this paper, we examine the galactic environment of metal-line systems,
a principal goal of the full survey.
These systems are of particular interest at present, because the 
\ion{O}{6}
transition provides the most efficient means of probing warm-hot 
$(T\sim 10^{5-7}$K) ionized gas in the intergalactic 
medium\footnote{At $T > 10^{5.5}$K, O$^{+6}$ and O$^{+7}$ are the
dominant ions in the WHIM \citep[e.g.][]{chenx03}, but these 
ions require X-ray observations and current instrumentation
limits one to only a few sightlines.}
\cite[the so-called WHIM;][]{tripp00,danforth05}.
Numerical simulations of the low redshift universe suggest the WHIM
is a major reservoir of baryons, which can be directly tested using
the \ion{O}{6} absorbers \citep{cen99,dave01b}.
This paper represents our group's first step 
toward understanding the physical
environment of low-redshift \ion{O}{6} systems.

Unless otherwise indicated, all distances in this paper are
physical separations derived from the
angular diameter distance, assuming a $\Lambda$CDM cosmology
with $\Omega_m = 0.3, \Lambda = 0.7, H_0=75 {\rm km s^{-1} Mpc^{-1}}$.
The physical separation is most meaningful for gravitationally
bound systems.  To obtain
transverse comoving separations, the physical separation can
be multiplied by (1+z).  All distances quoted scale with $h_{75}^{-1}$
with $h_{75} = H_0/75 {\rm km s^{-1} Mpc^{-1}}$.

\section{Photometry}

To perform target selection for multi-slit spectroscopy, we imaged 
the field surrounding PKS0405--123 with the Swope 40$''$ telescope
at Las Campanas Observatory.  Our first set of images were acquired on
UT Oct 26, 2000 with the SITe1 CCD in direct imaging mode
(pixel size $0.6964''$ for the 2048x2048 array).
These images were centered on PKS0405--123 and covered 
an $\approx 20' \times 20'$ field-of-view (FOV).  
We obtained $B$ and $R$ images under photometric (but windy) conditions
for total exposure times of 900s and 350s respectively.  We obtained
multiple exposures with a $10''$ dither patter to account for bad pixels
and to facilitate the construction of a super-sky flat.

The imaging data were reduced with standard IRAF tasks to subtract the
overscan and bias, and to flatten the images.  We then used a set of
custom
routines to determine integer offsets between the images and combine the 
individual frames weighting by signal-to-noise.  Finally, we derived
a photometric solution (zero point, airmass, and color terms) on the
Vega scale from
standard star observations taken throughout the night:
$B_{ZP} = 21.54 \pm 0.01; B_{AM} = 0.24 \pm 0.01; R_{ZP} = 21.81 \pm 0.01;
R_{AM} = 0.08 \pm 0.006; (B-R)= -0.03 \pm 0.002.$
The $B-R$ color term is sufficiently small that we choose to ignore it 
in all subsequent analysis.

%\begin{table}[ht]\footnotesize
\begin{table}[ht]\scriptsize
\begin{center}
\caption{{\sc OBJECT SUMMARY \label{tab:obj}}}
\begin{tabular}{rccccccc}
\tableline
\tableline
ID & RA & DEC & $R$ & S/G$^a$ & Area& flg$^b$ & $z$ \\
 & & & (mag) &  & ($\square''$) \\
\tableline
98&04:08:32.0&--12:17:07&$18.96 \pm 0.07$&0.87&   3.7& 7&$0.26016$\\
118&04:08:31.2&--12:21:21&$19.76 \pm 0.07$&0.88&   3.4& 7&$0.33226$\\
187&04:08:29.4&--12:16:55&$18.45 \pm 0.07$&0.02&   7.1& 7&$0.17736$\\
221&04:08:29.0&--12:13:28&$19.40 \pm 0.07$&0.14&   4.5& 7&$0.34660$\\
258&04:08:27.4&--12:22:01&$18.00 \pm 0.06$&0.51&   5.9& 7&$0.08193$\\
270&04:08:27.5&--12:16:03&$18.82 \pm 0.07$&0.18&   6.7& 7&$0.13360$\\
289&04:08:27.2&--12:14:12&$19.80 \pm 0.07$&0.14&   4.2& 7&$0.24012$\\
312&04:08:26.9&--12:12:10&$19.52 \pm 0.07$&0.81&   4.0& 7&$0.51655$\\
324&04:08:32.0&--12:07:09&$19.93 \pm 0.07$&0.47&   3.2& 7&$0.30934$\\
366&04:08:25.0&--12:16:36&$18.43 \pm 0.07$&0.97&   4.6& 7&$0.17706$\\
\tableline
\end{tabular}
\end{center}
\tablenotetext{a}{Star/galaxy classifier calculated by SExtractor.
Values near unity indicate \\
a stellar-like point-spread function.}
\tablenotetext{b}{This binary flag has the following code:
1: Photometry; 2: Spectrum \\
taken; 4= Redshift determined.}
\tablecomments{[The complete version of this table is in the electronic 
edition \\ 
of the Journal.  The printed edition contains only a sample.]}
\end{table}

Objects in the stacked $R$ image were identified using 
SExtractor \citep[v. 2.0][]{bertin02}.  
We required a minimum detection area of
6~pixels and a detection threshold of 1.5$\sigma$ above the sky RMS.
We constructed segmentation maps (images which consist of those pixels
associated with each galaxy identified by SExtractor)
and then calculated $B$ and $R$ magnitudes
adopting the photometric solution listed above and assigning the 
average airmass of the combined images.  We also used SExtractor to
calculate a star/galaxy classifier which is one criterion of the 
target selection algorithm.  
Table~\ref{tab:obj} lists the RA, DEC, $R$ magnitude and error,
the star/galaxy classifier, and the area in $\square''$ for each
object identified by SExtractor.  Note that a small fraction of
these objects may be spurious (e.g.\ bad pixels).

We acquired additional $R$ images of the field on the photometric
nights of UT 2002 Oct 3 and 4 to expand the field-of-view 
to $\approx 1 \square^\circ$.
Unfortunately, owing to an error in coordinates, the area imaged
is discontiguous.  For these observations, we used the SITe3 CCD in
direct imaging mode (pixel=0.435$''$ on this 2048x3150 array)
and acquired multiple dithered exposures at each pointing.

For this set of observations, the images were processed with a package
of IDL routines developed by JXP for direct 
images\footnote{Publically available 
http://www.ucolick.org/$\sim$xavier/IDL/index.html}.
The algorithms are similar to standard IRAF tasks yet are tailored to
this specific dataset.  The routines perform overscan subtraction,
the construction of a super-sky flat, and the flattening of science
exposures.  We also used the package to derive a photometric solution
for the two nights. 
Finally, offsets were determined for the individual exposures and
the images were combined to create a stacked image for each of four
pointings.

\clearpage

\begin{figure}[ht]
\begin{center}
\includegraphics[angle=90,width=3.5in]{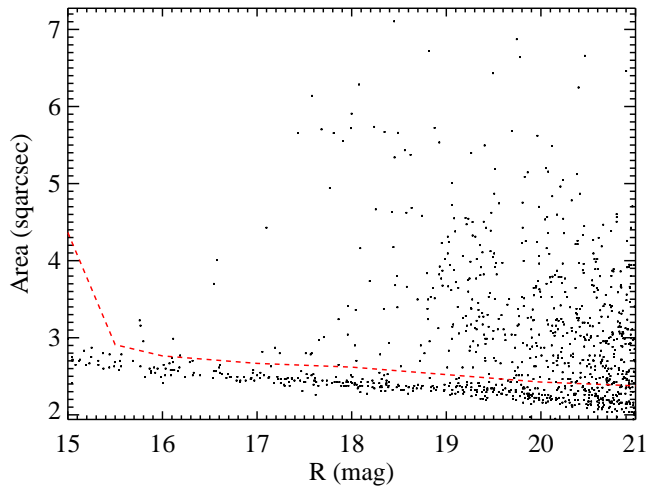}
\caption{Plot of the isophotal area (in square arcseconds) of
all the objects within $10'$ of PKS0405--123 as a function of 
$R$ magnitude.  The dashed line separates the targets we
pre-selects as stars and galaxies (stars are below the line).
Note that the stars follow a well-defined locus to $R\approx 19.5$\,mag.
}
\label{fig:area}
\end{center}
\end{figure}

\section{Target Selection and Mask Design}

The next task was to select objects from Table~\ref{tab:obj}
for follow-up spectroscopy with the WFCCD spectrograph on the 
100$''$ Dupont telescope at Las Campanas Observatory.  
Our goal was to achieve $> 90\%$
completeness to $R \approx 20$ within a circle of 10$'$ radius centered
on PKS0405--123 and the same completeness to $R \approx 19.5$ for
the flanking fields.  
For the central field, these observational parameters correspond
to a luminosity $L \approx 0.1 L_*$ and radial field-of-view of 1\,Mpc
at $z=0.1$.  
These magnitude limits represent a 
compromise between surveying down the luminosity function and
the sensitivity limit of the WFCCD spectrograph.
To distinguish stars from extragalactic sources, we examined
the area of the objects as a function of magnitude (Figure~\ref{fig:area}).
Point sources should occupy a uniform locus at the bottom edge of
the distribution and can be removed from the target list accordingly.
The line in Figure~\ref{fig:area} represents our area-cut as a 
function of magnitude;  objects below the line were discarded as
stars with the exception of those objects where the star/galaxy
classifier from SExtractor is less than 0.98.
This value was adopted by comparing the full set of values against
the results presented in Figure~\ref{fig:area}.
Although the seeing was poorer for the imaging of the flanking
fields, the smaller pixel size gives more reliable star/galaxy
separation.

There are two problems with this procedure:  (1) compact galaxies
may be misidentified as stars; (2) galaxies and stars are not well
separated at faint magnitudes.  We addressed the latter concern by 
taking a very conservative cut at faint magnitudes.
The former issue was investigated by
observing a mask of objects drawn from beneath the line
in Figure~\ref{fig:area}.  Of the 34 targets observed, all were
spectroscopically classified as stars.  
Nevertheless, we expect that our survey has
a few percent incompleteness due to this target selection criterion,
especially for objects fainter than $R=19.5$\,mag.
Finally, we inspected each target galaxy to remove any erroneous
objects from the SExtractor results.
In a handful of cases, we removed an object because of its proximity
to a very bright star which negatively affected its photometry and
isophotal area.

\begin{figure}[ht]
\begin{center}
\includegraphics[width=3.5in]{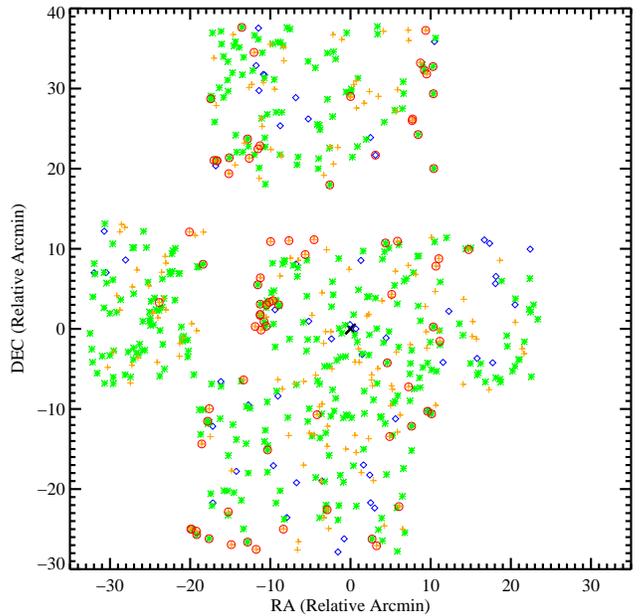}
\caption{Sky position of the galaxies in the field
surrounding PKS0405--123 (marked by the 'x' at RA=DEC=0).
The diamonds indicate galaxies brighter than $R=18$\,mag, the
asterisks are galaxies with $18<R<19.5$, and the 
plus-signs are for galaxies with $19.5 < R < 20$.
The circled objects do not have a measured redshift.
}
\label{fig:radec}
\end{center}
\end{figure}

In Figure~\ref{fig:radec} we show positions of all the target galaxies for
the center and flanking fields.  
We designed 24 masks targeting 549 objects in the full field.  The
masks were not designed such that the slits would be observed at
the parallactic angle, but 
we used $1.5''$ wide slits and losses due to 
atmospheric dispersion were not severe.
The various point styles in Figure~\ref{fig:radec} identify 
galaxies with several magnitude ranges
and the circles indicate objects with unknown redshifts.
Table~\ref{tab:obj} also lists the redshift of each object and the
galaxy type as defined in the next section.

\section{WFCCD Spectroscopy}

\subsection{Observations}

At the time we initiated this survey, the Dupont
Wide Field CCD camera \citep{weymann01}
was the largest field-of-view multi-slit 
spectrometer on any telescope.  
Its combination of field-of-view and throughput
were essential to surveying galaxies over a large area to a
sensitive luminosity limit.   All of our observations were 
performed with brass slit masks milled at the observatory with
slits covering an approximately $20' \times 15'$ field-of-view.
We employed the blue grism which has a dispersion of
FWHM~$\approx 375 \mkms$.

With few exceptions, we observed each mask for two exposures
totaling 1hr of integration.  In the first few years of operation,
the WFCCD spectrometer suffered from significant flexure.
Although the largest contribution to the flexure 
was fixed prior to our observations,  we chose to take a coeval
set of flats and arc-line calibration frames for each mask during
the night, interleaved with the observations.
The observations were carried out during two runs with the Dupont
Telescope in Sep 2001 and Oct/Nov 2002.  With little exception
the nights were photometric.
We used the TEK\#5 CCD in each case.  

For slits milled near the center of the mask, the spectral
coverage spans $\lambda = 3800 - 9000$\AA, with a spectral
dispersion of 2\AA~pix$^{-1}$ and a spatial scale of
$\approx 0.75''$~pix$^{-1}$.

\subsection{Data Reduction Pipeline}

All of the spectroscopic data were reduced and analyzed with
an IDL package specifically designed for the WFCCD instrument.
This subsection serves as the official release of the WFCCD reduction
package\footnote{http://www.ociw.edu/lco/dupont/instruments/manuals/wfccd/redux}
and we will describe at greater length its procedures
(for recent modifications to the package, see Geha \& Blanton 2005 in prep.).
Given a night of data, the user creates an IDL structure which 
describes the key diagnostics of each image (e.g.\ filter, exposure time,
UT, image type, mask ID).  A fraction of this information is
parsed from the relatively sparse FITS header of WFCCD data and
the remainder is input by the observer.  
After this initial step, the pipeline is fairly automated and
is guided by this data structure.

The steps of the data reduction for a given mask are as follows.
The dome flat exposures are scaled and combined and cosmic rays are
conservatively identified in cases with multiple exposures.
A sawtooth image of the flat is generated by differencing the image
with itself after shifting it by a single row (spectra run parallel to the 
CCD rows).
The resulting image exhibits a sawtooth pattern at each slit edge.
Each slit edge then shows a sharp feature which is traced to determine
the slit curvature and to construct a two-dimensional map of the
y-distortion.  
The y-distortion map is used to rectify (in the spatial direction only
with the sole goal of mapping the wavelength solution as noted below)
the flat image and the
slit edges are cross-correlated with the expected slit positions
(determined from the output of the mask design software) to
determine the true position of each slit.  Finally, the flat
for each slit is normalized by the median of the central five rows
in the original (i.e.\ curved) frame and a final image for 
the normalization of pixel-to-pixel variations is created.

The pipeline then overscan subtracts and flattens the arc images
associated with the mask and rectifies the frames.  A one-dimensional
arc spectrum is extracted down the middle of each slit and a wavelength
solution is automatically derived using the known spatial coordinates
of the slit as input.  The dispersion of this grism spectrograph
is non-linear and a 6th-order polynomial is required to achieve
a fit with a typical RMS of 0.2~pixels.  The pipeline then traces
the curvature of the arc lines within each slit and fits a low-order 
polynomial to each line.  It then generates a unique wavelength
solution for each row in the slit, i.e.\ every pixel in the
rectified arc image is assigned a unique wavelength.  Lastly, this
wavelength image is mapped back to the original data frame using
the inverse of the y-distortion map.

The remainder of the data reduction is carried out on the science
images without rectification.  Each exposure is overscan subtracted
and flattened, a variance array is constructed assuming Poisson
statistics and the read noise (5.6 electrons for gain=1),
and the wavelength image of the arc closest in UT is assigned to the
exposure.  If two or more science exposures were obtained (without
dithering along the slits), cosmic rays are masked using a conservative
algorithm to avoid mis-identifying sharp emission line features.
The pipeline then identifies all objects with significant flux in
each slit.  By default, the algorithm searches for peaks in the flux
along the slit at $\lambda = 5500$\AA.  If a peak is identified
within 5~pixels of the expected position of the science object, it
is assigned as the primary object.  If no object is identified within
5~pixels of the expected position, the program gives the primary
object a centroid corresponding to the expected position.
Any additional objects are flagged as serendipitous. 
The code then masks the regions surrounding all identified objects and
the centroids will be used as starting points for tracing and
extraction.
A GUI is launched to allow the user to verify object identification
and to modify the regions of the slit used to perform sky subtraction.
The object centroids are saved within an IDL structure 
for each science exposure.

\begin{figure}[ht]
\begin{center}
\includegraphics[angle=90,width=3.5in]{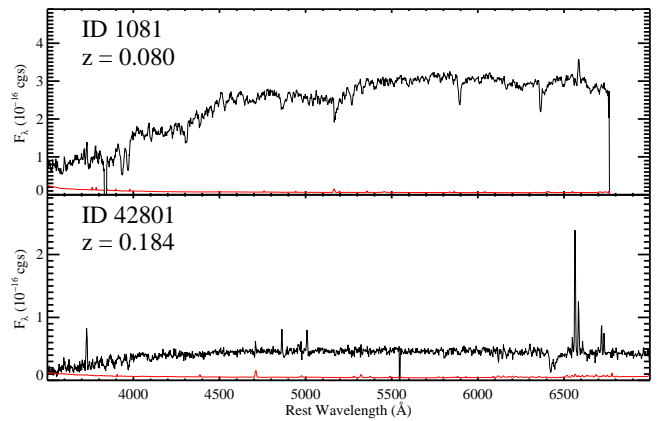}
\caption{Example galaxy spectra for two of the brighter
galaxies in the survey.  The drop-off in flux of galaxy 1081
at $\lambda \approx 6770$\AA\ is due to the CCD edge.
Also, the feature at $\lambda \approx 3820$\AA\ is due to 
a series of bad columns on the CCD.  The red line traces the
1$\sigma$ error array.
}
\label{fig:galex}
\end{center}
\end{figure}

Sky subtraction is performed by fitting those pixels 
within a given slit that were not
masked as object pixels.  
The wavelength and flux values for each unmasked pixel in
the slit are used to estimate the sky spectrum.  The values are
fit with a third order b-spline with breakpoints centered
on every other column. 
The b-spline is then evaluated at every pixel in the slit and subtracted.
The routine iterates over all slits and the full sky subtracted image
is appended to the processed data FITS file.  
Finally, the objects are extracted using boxcar extraction.
The spatial profile is fit with a spline and an aperture containing
95$\%$ of the flux is defined.  This profile is also used to flag
cosmic rays within the extraction window.  These pixels are flagged and
their corresponding 1D spectral regions are masked (e.g.\ given zero
weight).
The 1D spectra of individual exposures is scaled and combined with
optimal weighting.  The routine flags and rejects additional bad pixels
for masks with two or more coeval exposures. 
The final 1D spectra are fluxed with a fixed sensitivity function.
The sensitivity function is valid for the majority of
slits but does not account for vignetting at the edges of the CCD.
Figure~\ref{fig:galex} presents the spectra of two relatively bright
galaxies from our sample.  All of the spectra and fits tables
are available at http://www.ucolick.org/$\sim$xavier/WFCCDOVI/.

\subsection{Galaxy Redshifts, Completeness, and Selection Functions}

An initial redshift for every extracted object was calculated 
with an automated procedure based on the Sloan Digital Sky
Survey (SDSS) software routine {\it zfind}.  We first smoothed and
rebinned the four SDSS galaxy and star
eigenfunctions (EigenGal-52223, EigenStar-52374) to the WFCCD
resolution and wavelength range.  We then implemented a slightly
modified version of {\it zfind} which allows for greater 
freedom in the continuum of the galaxy due to poor fluxing of
the WFCCD data (in particular those galaxies grossly
affected by vignetting).
The algorithm steps from $z=0.05$ to 0.5 in steps of the extracted
pixel width (100 km/s).  At each step it fits a linear combination
of the galaxy eigenfunctions to the pixel data and variance 
by minimizing $\chi^2$.  It then finds the minimum of the array
of $\chi^2(z)$ and fits a curve to the array near the minimum 
to yield a redshift with formal 1 sigma 
uncertainty of $\sim 30$\kms\ for the majority of spectra.
We expect that systematic uncertainty
(e.g.\ limitations of the eigenspectra)
are smaller than the statistical error. 
Finally, the spectrum and best-fit were visually inspected
by two authors and the redshift was corrected for $\approx 10\%$
of the objects.  In these cases, we reapplied the automated
algorithm to find the minimum $\chi^2$ in a restricted redshift 
interval.
The redshift values are listed in column 8 of Table~\ref{tab:obj}.

\begin{figure}[ht]
\begin{center}
\includegraphics[angle=90,width=3.5in]{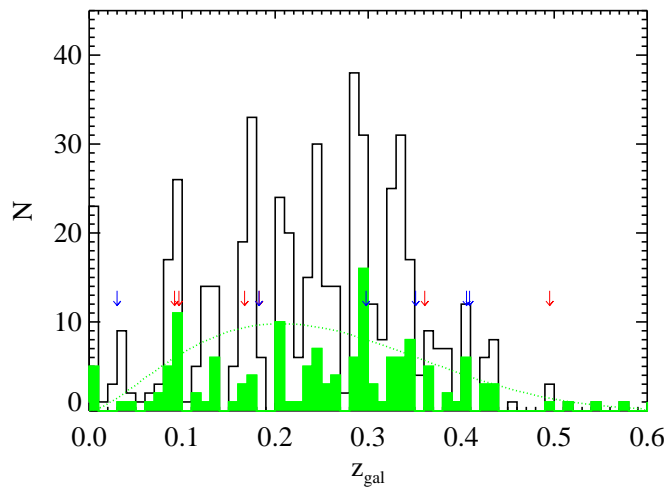}
\caption{Histogram of the galaxy redshifts for the field surrounding
PKS0405--123.  The solid histogram shows the results for galaxies
within $10'$ of PKS0405--123 while the open histogram shows the
distribution for our full sample.  The red arrows indicate the
redshifts of metal-line systems along the sightline.  The blue
arrows denote absorption systems with $\mnhi > 10^{14} \cm{-2}$ but
no apparent metal absorption.  The objects in the bin at $z=0$ are stars.
}
\label{fig:zhist}
\end{center}
\end{figure}

Figure~\ref{fig:zhist} presents histograms of the redshifts
for all of the primary objects in the sample.
The solid bars are the histograms for galaxies within 
$10'$ of PKS0405--123 and the open histogram shows the full sample.
Overplotted on the histograms is the selection functions
for the survey (i.e.\ the predicted redshift distribution based
on the survey limit;  see Paper~II for details).  
%For the inner sample, we calculated the 
%sensitivity function by adopting the luminosity function of
%\cite{blanton01} 
%and a survey limit of $R=20$\,mag with 95$\%$ completeness.
%For the full sample, we adopted a magnitude limit $R=19.5$\,mag
%for the flanking fields (covering $\approx 1 \square^\circ$) and a survey
%completeness of $\approx 90\%$.

\begin{figure}[ht]
\begin{center}
\includegraphics[angle=90,width=3.5in]{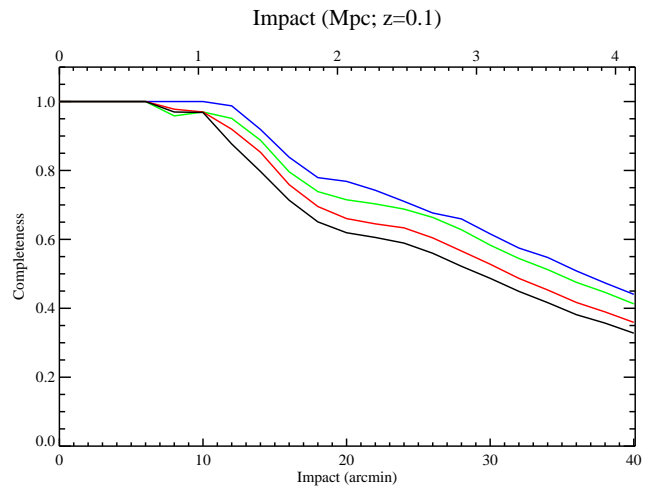}
\caption{Completeness curves for the galaxy survey as a function
of impact parameter to the quasar.  
The curves refer to limiting 
magnitudes $R_{lim} = 18, 19, 19.5, 20$\,mag. 
The measurement is a direct
evaluation for galaxies within $10'$ of the quasar.  
At radii greater than 10$'$, the incompleteness is primarily
due to the discontiguous coverage of the flanking fields
(Figure~\ref{fig:radec}).
At these radii, we used the surface density of 
objects within $10'$ to estimate
the incompleteness in the discontiguous regions.
}
\label{fig:complete}
\end{center}
\end{figure}

In Figure~\ref{fig:complete}, we report the completeness of the
survey as a function of impact parameter for several magnitude
limits.  For the inner $10'$ the curves simply indicate the fraction
of galaxies with measured redshifts that are
brighter than the magnitude limit.
At larger impact parameter, we use the observed surface density
within the inner $10'$ to predict the number of galaxies and
calculate the ratio of galaxies with measured redshift versus
the predicted number of galaxies.
Note that none of the calculations account for a incompleteness
due to our star/galaxy algorithm.
We believe we are overestimating the completeness by a few percent
for $R > 19.5$\,mag.

\begin{figure}[ht]
\begin{center}
\includegraphics[angle=90,width=3.5in]{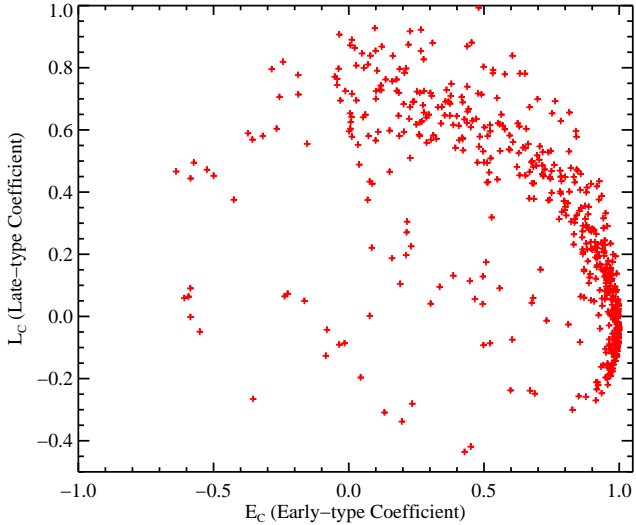}
\caption{Scatter plot of the early and late-type coefficients,
defined as described in the text.  With a few exceptions, the
galaxies follow a well-defined locus.  The majority of outliers
are either quasars or galaxies at very high redshift where we have
limited data for spectral typing.
}
\label{fig:eigen}
\end{center}
\end{figure}

\subsection{Galaxy Types}

The eigenfunction coefficients (which are independent of galaxy brightness)
from the best-fit redshift solution
can be used to crudely assign a galaxy spectral type.
Although the four SDSS galaxy eigenfunctions are non-physical
(e.g.\ functions three and four include `negative' emission lines),
the first eigenfunction is dominated by absorption features
and the second eigenfunction has significant H$\alpha$,
[\ion{O}{2}], and other strong emission lines.  Therefore,
we defined an `early-type' coefficient $E_C$ equal to the eigenvalue
for the first eigenfunction (maximum value of 1)
and a `late-type' coefficient $L_C$
which is the second eigenvalue minus the sum of the third and
fourth coefficients.  
This definition was determined by examining
the eigenvalues for a subset of the emission-line galaxies.
The functions described by these two
coefficients are orthogonal but it is possible for a galaxy
spectrum to have non-negligible values for both $E_C$ and $L_C$.
Figure~\ref{fig:eigen} presents the values for all of the galaxies
observed in the PKS0405$-$123 field.  In general, the galaxies lie
along a well-defined locus although there are notable exceptions.
These are primarily quasars or high $z$ galaxies with minimal 
coverage of the key spectroscopic features.
In the following, we will distinguish early-type galaxies 
as galaxies with $E_C > 0.8$ and $L_C < 0.4$ and late-type
galaxies with $L_C > 0.4$ and $E_C < 0.8$.

\section{RESULTS AND DISCUSSION}

In this section we discuss the results of our galaxy survey
in terms of the origin of low $z$ metal-line systems.
For each known metal-line absorber, we examine the luminosity range,
projected distances, and number density of the surrounding galaxies 
over a large volume to the magnitude limit defined in our redshift 
survey.  This is different but complementary to the approach adopted by 
\cite{chen01}, who investigated the presence/absence of
metal absorbers for each known galaxy at small projected distance
to the sightline.

\subsection{Individual Metal-Line Systems}

We begin with a discussion of the galaxies
associated with each of the metal-line systems identified along
the sightline to PKS0405--123 (Paper~I).  We focus our attention on the
galaxies with smallest impact parameters and also any large-scale
structures identified in our survey.  Table~\ref{tab:mtlglx} lists
the properties of the galaxies identified with $|\delta v| < 1000 \mkms$
where

\begin{equation}
\delta v \equiv c \frac{(z_{abs}-z_{gal})}{(1+z_{abs})} \perd
\end{equation}

\noindent  This velocity limit is somewhat arbitrary.
We chose the value because it characterizes the peculiar
velocities of galaxies in the largest, gravitationally
bound structures.  We will, however, primarily focus on
galaxies with $|\delta v| < 500 \mkms$ which are more
likely to be physically associated with the absorption systems.
We do not impose a limit to the impact parameter for this
discussion, in part because we wish to consider the role
of large-scale structures (e.g.\ filaments).

Figure~\ref{fig:galx} presents a visual survey of the galactic
environment of each absorber.
For the metal-line systems, our galaxy survey covers from
$\approx 3$ to 10 physical Mpc in radius for $z = 0.1$ to 0.4.
Therefore, the survey is not sensitive to very large-scale
structures at $z \sim 0.1$, although we emphasize that environmental
correlations with galactic properties appear to be dominated
by the environment measured on 1 to 3\,Mpc scales 
\citep[e.g.][]{blanton05}.

Table~\ref{tab:mtlglx} also summarizes the physical properties
of the metal-line absorbers, each of which has
$\mnhi > 10^{14} \cm{-2}$.  Furthermore,
all of the systems show \ion{O}{6} absorption except the absorber at
$z=0.3608$ where an \ion{O}{6} absorber is 
identified\footnote{\cite{williger05} also
report the detection of an additional \ion{O}{6} system at 
$z=0.3616$ ($\delta v = 240 \mkms$).  
We are not convinced of the identification based on
our extraction of the STIS data and do not consider it here.  In any case,
its existence does not bear significantly on our results.}
nearby ($\delta v = 550 \mkms$).
These \ion{O}{6} systems are 
particularly notable because they provide the most efficient
means of probing the warm-hot intergalactic medium 
\citep[WHIM;][]{tripp00} with current instrumentation.
Numerical simulations suggest that the WHIM is a major baryonic
component of the low redshift universe \citep{cen99,dave01b}.  
They describe the WHIM as a low density gas with temperature
$T \approx 10^{5-7}$K which has been shock heated during the
formation of large-scale filamentary structures.
If this gas is metal-enriched, then it would be detectable
with UV spectroscopy via the \ion{O}{6} transition.

\subsubsection{$z=0.0918$}

\begin{figure*}[ht]
\begin{center}
\includegraphics[angle=90,width=6.5in]{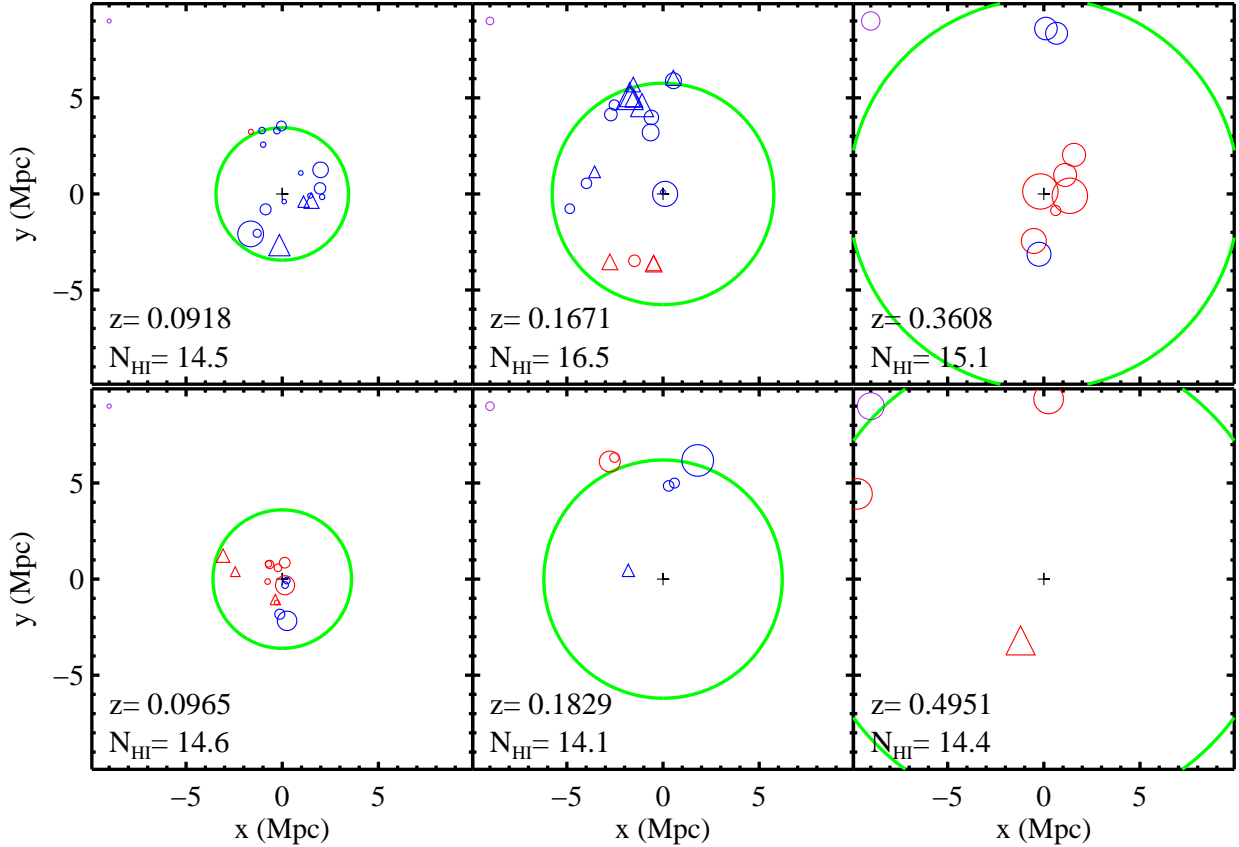}
\caption{Plot of the spatial distribution of galaxies around
PKS0405--123 for each of the metal-line absorption systems detected
along the sightline.  
Galaxies are color-coded according to their velocity offset from the
absorption system and circles indicate $|\delta v| < 500\mkms$
and triangles are for $1000 > |\delta v| > 500 \mkms$.
The size of the symbol is proportional to the
luminosity of the galaxy.
The purple circle
in the upper-left corner of each sub-panel corresponds to 
$R=20$\,mag, i.e. the magnitude limit of the inner $10'$ of the survey.
This corresponds to $L \approx L_*/10$ at $z=0.1$ and 
$L \approx L_*$ at $z=0.25$.
The green circle in each sub-plot shows the
approximate radial limit of the survey ($36'$).  
}
\label{fig:galx}
\end{center}
\end{figure*}

Of all the metal-line absorbers identified toward PKS0405--123,
the \ion{O}{6} system at $z=0.0918$ shows the largest number of
galaxies with $|\delta v| < 500 \mkms$. 
It is also the lowest redshift absorber
and, therefore, we probe furthest down the luminosity function.
Although one identifies 14 galaxies,
only one of
these has an impact parameter $\rho < 1$\,Mpc and even it
lies at $\rho \approx 400$\,kpc.  This low luminosity, emission-line
galaxy has a velocity offset $\delta v = -340 \mkms$ 
which is much larger than the virial velocity of the galaxy and
a physical association is unlikely.
There are no additional galaxies to within 1\,Mpc of the quasar
sightline and we emphasize that the survey completeness is 
greater than $95\%$ to $R=20$ ($L \approx L_*/10$) at this impact parameter.
It is unlikely, then, that this absorber is 
physically associated with a galactic halo.
Examining Figure~\ref{fig:galx}, one identifies three `groups' of galaxies
at $z \approx 0.091$ roughly surrounding the quasar sightline.
The overall impression is that this gas occurs near
the intersection
of several groups, perhaps related to an over-dense, large-scale 
structure (i.e.\ filament).

\subsubsection{$z=0.0965$}

%In contrast with the $z=0.0918$ absorber, the metal-line system
%at $z=0.096$ has 7~galaxies with $\rho \lesssim 1$\,Mpc and the
%closest have velocity offsets $|\delta v| < 60 \mkms$.
%The two closest galaxies are late-type systems and, given their
%low luminosity, it is unlikely the absorption system lies within
%their virial radius.  The gas may lie in the halo of the
%bright, early-type galaxy (ID 1601) at impact parameter $\rho = 350$\,kpc.
%Although we cannot rule out the possibility that the absorber
%[resides] within one of these galactic halos, the large impact
%parameters argue the systems is associate with intragroup gas.
%This galaxy group contains several $L \gtrsim L_*$ galaxies, including three
%early-type galaxies.  
%The galaxies with $|\delta v|<500 \mkms$ surround the sightline
%suggesting the gas is physically bound to the group.

A strong case can made that this \ion{O}{6} absorber arises in an
intragroup medium.   First, the sightline penetrates a galaxy
group containing several $L \gtrsim L_*$ galaxies, including three
early-type galaxies.  Restricting the analysis to the galaxies within
2.2\,Mpc, the group has a velocity centroid offset of
$140\mkms$ from the gas and a velocity dispersion of 210\kms.
Second, the galaxies with $|\delta v|<500 \mkms$ surround the sightline
and the gas must be gravitationally bound to the group.
Third, although there are three galaxies with $\rho = 250-350$\,kpc, 
these have too low luminosity for their virial radius to encompass the
sightline. 
In short, it is more likely the intragroup 
medium gives rise to this \ion{O}{6} 
absorption system \citep{mulchaey96,fukugita98}.
In Paper~I we estimated the volume density for this gas 
$n_H \approx  10^{-5} \cm{-3}$
assuming the extragalactic ultraviolet background radiation field from
\cite{dave01}.
The gas density profile of the intragroup medium can be
estimated directly for X-ray detected groups from the X-ray 
surface brightness profile.
For an X-ray group with central volume density  $n_H \approx 10^{-3} \cm{-3}$
and a standard beta model 
\citep[i.e.][]{helsdon00,mulchaey03},
%(i.e. Helsdon & Ponman 2000; Mulchaey et al. 2003),
the density at $\rho = 100$\,kpc 
is $\approx 10^{-4} \cm{-3}$. 
If the group at $z=0.0965$ is not an X-ray group,
one might expect a significantly lower gas density, 
i.e. one consistent with $10^{-5} \cm{-3}$ \citep{mulchaey96}.
We also note, however, that this system exhibits \ion{C}{3}
absorption and we have argued that the gas is predominantly photoionized
and in a single phase (Paper~I).  
We cannot rule out, however, that the 
the \ion{O}{6} gas is in a multi-phase medium. 

\subsubsection{$z=0.16710$}

This partial Lyman limit system has been previously identified with
two galaxies at $\rho \approx 100$\,kpc \citep{spinrad93}.  
One of these is a bright early-type galaxy and the other is a 
dwarf galaxy with modest emission lines.
These are the only two galaxies in our survey of the PKS0405--123
field with impact parameter less than 150\,kpc.
Prior to performing
our survey, we expected to identify additional galaxies at the redshift
of the absorber in the field surrounding the sightline.
To our surprise, we find no additional galaxies at $z \approx 0.167$
with $L > 0.5 L_*$ within 3\,Mpc of the absorber.  In fact, there
is only one additional galaxy with $|\delta v| < 300 \mkms$ and it lies
at an impact parameter of nearly 4\,Mpc.  
%Granted the updated survey, we identify the absorber with the
%bright galaxy at $\approx 100$\,kpc impact parameter.
We will return to this absorber in $\S$~\ref{sec:1671}.

\clearpage

\begin{table*}[ht]\footnotesize
\begin{center}
\caption{{\sc SUMMARY OF THE GALAXIES NEIGHBORING METAL-LINE ABSORPTION SYSTEMS\label{tab:mtlglx}}}
\begin{tabular}{rcccccccccc}
\tableline
\tableline
ID & z & R & Lum & $\delta v$ & $\rho$ 
& E$_c$ & L$_c$ \\
 & & & ($L^*$) & (km/s) & (kpc) \\
\tableline
\tableline
$\,\;\;\;\;\;\;\;\;\;\;$& 
$\,\;\;\;\;\;\;\;\;$& 
$\,\;\;\;\;\;\;\;$& 
$\,\;\;\;\;\;$& 
$\,\;\;\;\;\;$& 
$\,\;\;\;\;\;$& 
$\,\;\;\;\;\;\;\;$& 
$\,\;\;\;\;\;\;\;$&  \\
\end{tabular}

%\cutinhead{$z=0.0918, \mnhi=14.5, \N{OVI} = 13.8$, [M/H] = $>-1.4$}
$z=0.0918, \mnhi=14.5, \N{OVI} = 13.8$, [M/H] = $>-1.4$ \\
\begin{tabular}{rcccccccccc}
\tableline
1698& 0.0906& 19.74&  0.13&  -341&   412.4&  0.49&  0.04& \\
10258& 0.0899& 17.72&  0.87&  -517&  1175.1& -0.03&  0.69& \\
3386& 0.0917& 17.81&  0.80&   -24&  1181.2&  1.00& -0.03& \\
10552& 0.0916& 19.52&  0.17&   -60&  1455.3&  0.98&  0.20& \\
11489& 0.0903& 19.74&  0.14&  -403&  1457.6&  0.60& -0.24& \\
10309& 0.0899& 17.09&  1.56&  -531&  1552.4& -0.52&  0.47& \\
10900& 0.0912& 17.74&  0.85&  -161&  1986.0&  0.05&  0.64& \\
10498& 0.0908& 19.35&  0.19&  -276&  2081.1&  0.95&  0.07& \\
11603& 0.0908& 17.06&  1.59&  -278&  2354.3&  0.97&  0.14& \\
30847& 0.0902& 18.49&  0.43&  -436&  2427.6& -0.24&  0.06& \\
30799& 0.0917& 15.97&  4.37&   -21&  2650.3&  0.94& -0.16& \\
30260& 0.0899& 16.38&  2.99&  -522&  2669.7&  0.99& -0.08& \\
41345& 0.0906& 19.37&  0.19&  -333&  2748.4& -0.02& -0.09& \\
42607& 0.0911& 18.94&  0.28&  -206&  3305.6&  0.56&  0.09& \\
42613& 0.0908& 18.97&  0.27&  -263&  3462.5&  0.94&  0.39& \\
43005& 0.0910& 18.08&  0.62&  -226&  3536.0&  0.40&  0.64& \\
42509& 0.0924& 19.57&  0.16&   153&  3623.5&  0.50&  0.13& \\
\tableline
\end{tabular}

%\cutinhead{$z=0.0965, \mnhi=14.7, \N{OVI} = 13.7$, [M/H] = $-1.5$}
$z=0.0965, \mnhi=14.7, \N{OVI} = 13.7$, [M/H] = $-1.5$ \\
\begin{tabular}{rcccccccccc}
\tableline
1457& 0.0965& 19.03&  0.29&   -13&   247.5&  0.18&  0.76& \\
1602& 0.0965& 19.01&  0.29&    -4&   342.7&  0.21&  0.27& \\
1601& 0.0967& 16.74&  2.39&    56&   350.7&  0.68&  0.38& \\
2254& 0.0970& 18.74&  0.38&   149&   629.9&  0.67&  0.46& \\
3138& 0.0980& 19.38&  0.21&   414&   770.8&  0.50& -0.09& \\
1659& 0.0970& 17.99&  0.75&   132&   864.8&  0.00&  0.65& \\
2973& 0.0971& 18.52&  0.46&   161&  1006.3&  0.48&  0.49& \\
3082& 0.0973& 19.06&  0.28&   228&  1068.0&  0.87&  0.36& \\
2485& 0.0986& 18.05&  0.72&   573&  1123.1&  0.36&  0.69& \\
32180& 0.0974& 19.77&  0.15&   244&  1250.0&  0.19&  0.58& \\
31271& 0.0957& 18.13&  0.66&  -222&  1832.6& -0.50&  0.45& \\
30806& 0.0964& 16.69&  2.50&   -18&  2186.3&  0.91&  0.05& \\
21111& 0.0994& 18.20&  0.62&   797&  2474.9&  0.88&  0.40& \\
21834& 0.1000& 17.49&  1.20&   970&  3307.2&  0.95&  0.15& \\
\tableline
\end{tabular}

%\cutinhead{$z=0.1671, \mnhi=16.5, \N{OVI} = 14.8$, [M/H] = $-0.2$}
$z=0.1671, \mnhi=16.5, \N{OVI} = 14.8$, [M/H] = $-0.2$ \\
\begin{tabular}{rcccccccccc}
\tableline
90033& 0.1670& 21.00&  0.15&   -26&    92.6& ...&...&\\
1753& 0.1670& 17.43&  4.14&   -15&   107.8&  0.65&  0.45& \\
40291& 0.1659& 18.30&  1.86&  -317&  3264.4&  1.00&  0.04& \\
30714& 0.1710& 18.49&  1.56&   997&  3629.7&  0.93& -0.04& \\
30693& 0.1708& 18.32&  1.83&   960&  3661.2&  0.07&  0.81& \\
21409& 0.1650& 19.08&  0.90&  -547&  3739.0&  0.70&  0.50& \\
30811& 0.1673& 19.11&  0.88&    48&  3782.8&  0.19&  0.72& \\
21074& 0.1652& 19.33&  0.72&  -488&  4019.2&  0.31&  0.88& \\
41036& 0.1652& 18.63&  1.37&  -497&  4025.4&  1.00& -0.02& \\
30761& 0.1698& 18.42&  1.66&   687&  4487.2&  0.25&  0.72& \\
41676& 0.1649& 17.61&  3.52&  -565&  4747.1&  0.39&  0.62& \\
20274& 0.1660& 19.52&  0.60&  -292&  4910.1&  0.51&  0.58& \\
41188& 0.1652& 18.93&  1.04&  -478&  4942.9&  0.92&  0.42& \\
42077& 0.1646& 18.15&  2.14&  -652&  5207.7&  1.00&  0.02& \\
41685& 0.1652& 19.36&  0.70&  -484&  5272.7&  0.24&  0.69& \\
42162& 0.1647& 17.88&  2.74&  -622&  5354.5&  0.97& -0.10& \\
42163& 0.1647& 17.28&  4.78&  -629&  5375.1&  0.95&  0.11& \\
42816& 0.1650& 18.59&  1.42&  -548&  5899.1&  0.68&  0.48& \\
42990& 0.1653& 18.42&  1.66&  -468&  5914.8&  0.16&  0.84& \\
43154& 0.1650& 18.52&  1.52&  -536&  6064.1&  0.29&  0.64& \\
\tableline
\end{tabular}

%\cutinhead{$z=0.1829, \mnhi=14.1, \N{OVI} = 14.0$, [M/H] = $>-2.0$}
$z=0.1829, \mnhi=14.1, \N{OVI} = 14.0$, [M/H] $>-2.0$ \\
\begin{tabular}{rcccccccccc}
\tableline
3623& 0.1796& 19.17&  1.02&  -830&  1863.9& -0.29&  0.80& \\
41580& 0.1819& 19.54&  0.73&  -261&  4859.0&  0.42&  0.69& \\
41714& 0.1823& 19.67&  0.65&  -155&  5032.4&  0.19&  0.10& \\
42851& 0.1822& 17.14&  6.61&  -176&  6432.7&  0.80&  0.34& \\
42801& 0.1842& 18.04&  2.87&   326&  6715.2& -0.19&  0.71& \\
42983& 0.1834& 19.74&  0.60&   121&  6811.1&  0.68&  0.04& \\
\tableline
\end{tabular}
\end{center}
\tablecomments{The galaxy summary is restricted to those galaxies within 1000 \kms\ of the absorption system. 
The \nhi, $\N{OVI}$, and [M/H] values are taken from Paper~I. 
The impact parameter refers to physical separation, not comoving.
}
\end{table*}

\clearpage

\begin{table}[ht]\scriptsize
\begin{center}
{\sc TABLE 2 CONTINUED} \\

\begin{tabular}{rcccccccccc}
\tableline
\tableline
ID & z & R & Lum & $\delta v$ & $\rho$ 
& E$_c$ & L$_c$ \\
 & & & ($L^*$) & (km/s) & (kpc) \\
\tableline
\tableline
$\,\;\;\;\;\;\;\;\;\;\;$& 
$\,\;\;\;\;\;\;\;\;$& 
$\,\;\;\;\;\;\;\;$& 
$\,\;\;\;\;\;$& 
$\,\;\;\;\;\;$& 
$\,\;\;\;\;\;$& 
$\,\;\;\;\;\;\;\;$& 
$\,\;\;\;\;\;\;\;$&  \\
\end{tabular}

%\cutinhead{$z=0.3608, \mnhi=15.1, \N{OVI} = <13.3$, [M/H] = $>-0.7$}
$z=0.3608, \mnhi=15.1, \N{OVI} <13.3$, [M/H] $>-0.7$ \\
\begin{tabular}{rcccccccccc}
\tableline
1967& 0.3612& 18.58&  8.24&    97&   220.5&  0.98&  0.18& \\
90068& 0.3610& 21.30&  0.67&    44&  1062.4& ...&...&\\
1039& 0.3610& 18.58&  8.27&    47&  1356.6&  0.98&  0.09& \\
1236& 0.3615& 19.52&  3.47&   156&  1482.2&  0.98&  0.02& \\
2170& 0.3609& 19.34&  4.11&    27&  2503.3&  0.99& -0.08& \\
907& 0.3617& 19.51&  3.50&   194&  2567.9&  0.63&  0.53& \\
32334& 0.3598& 19.43&  3.76&  -219&  3147.8&  0.99& -0.02& \\
41821& 0.3587& 19.63&  3.14&  -457&  8383.7&  0.98& -0.07& \\
41971& 0.3598& 19.56&  3.35&  -218&  8606.8&  0.99& -0.10& \\
\tableline
\end{tabular}

%\cutinhead{$z=0.4951, \mnhi=14.4, \N{OVI} = 14.3$, [M/H] = $>-0.3$}
$z=0.4951, \mnhi=14.4, \N{OVI} = 14.3$, [M/H] $>-0.3$ \\
\begin{tabular}{rcccccccccc}
\tableline
2496& 0.4992& 19.82&  5.56&   816&  3412.7&  0.94&  0.23& \\
41479& 0.4961& 19.78&  5.76&   201&  9377.7&  0.99&  0.01& \\
21907& 0.4953& 19.71&  6.17&    36& 10679.6&  0.98&  0.10& \\
\tableline
\end{tabular}
\end{center}
\tablecomments{The galaxy summary is restricted to those galaxies within \\
1000 \kms\ of the absorption system. The \nhi, $\N{OVI}$, and [M/H] \\
values are taken from Paper~I. The impact parameter refers to physical \\
separation, not comoving.
}
\end{table}

\subsubsection{$z=0.1829$}

There are two \lya\ absorbers at $z=0.1825$ and $z=0.1829$ 
\citep{williger05} with \ion{O}{6}/\ion{H}{1} ratios differing
by more than an order of magnitude (Paper~I).  Focusing on the
\ion{O}{6} absorber at $z=0.1829$, we identify zero $L \approx L_*$
galaxies within 2\,Mpc of the sightline and none within 5\,Mpc
with $|\delta v| < 800 \mkms$.  Although our survey limit
does not preclude the presence of a sub-$L_*$ galaxy, our
results show that there is no significant galaxy group or large-scale
structure associated with the gas.  This observation is qualitatively
different from all of the other metal-line systems toward
PKS0405--123.

\subsubsection{Systems with $z \approx 0.36$}

There are a pair of metal-line systems
at $z=0.3608$ and $z=0.3633$ with a velocity separation of
$\delta v = 550 \mkms$.
The absorber at $z=0.3608$ is notable for exhibiting several oxygen
ions yet no significant \ion{O}{6} absorption.  In contrast, the
$z=0.3633$ system exhibits \ion{O}{6} and likely \ion{C}{3} gas.
Our galaxy survey reveals a group of very bright, early-type
galaxies coincident with the $z=0.3608$ absorber.  
One may place the gas within the elliptical at $\rho = 220$\,kpc
(ID 1967) but more likely the gas is associated with the
intragroup medium of these galaxies or a galaxy below
our magnitude limit (i.e.\ $L < 2L_*$). 
In any case, this is a striking example of an over-dense
galactic environment hosting a metal-line absorption system.

\subsubsection{$z=0.4951$}

This strong \ion{O}{6} absorption system is at too high redshift
to even qualitatively discuss its galactic environment based on our current
survey.  In passing, we note two galaxies with $|\delta v| \lesssim 200 \mkms$
and $L \approx 6 L_*$ which are
at very large impact parameter ($\rho \approx 10$\,Mpc) 
from the quasar sightline.
We also note that the \cite{williger05} compilation reveals no 
additional galaxies at fainter magnitudes closer to the sightline.

\begin{figure}[ht]
\begin{center}
\includegraphics[angle=90,width=3.5in]{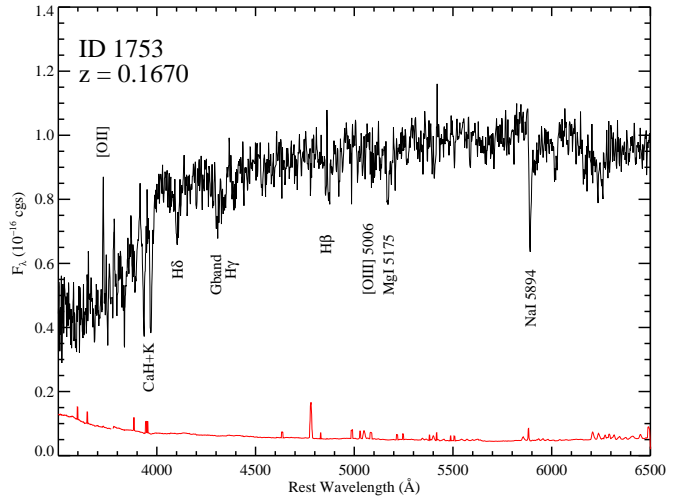}
\caption{Galaxy spectrum of the likely host galaxy of the
partial Lyman-limit absorption system at $z=0.1671$ toward
PKS0405--123.  The Balmer features, weak [OII] emission, and
absence of [\ion{O}{3}] emission mark this galaxy as an Sb spectral type.
This indicates minimal current star formation yet a significant
episode of star formation over the past $\sim 1$Gyr.
}
\label{fig:1671}
\end{center}
\end{figure}

\vskip 1.5in

\subsection{Halo Gas at Low Redshift}
\label{sec:1671}

Consider further the absorption system at $z=0.1671$. 
Combining our galaxy survey with the sample from 
\cite{williger05}, it is apparent the system is 
associated with one of the two galaxies at
impact parameter $\rho \approx 100$\,kpc from the sightline.
We consider it very unlikely that the gas is within the halo
of the low luminosity galaxy because (1) the gas has 
nearly solar metallicity and
(2) the gas is almost certainly gravitationally bound to the
bright spiral given the negligible velocity separation and 
large stellar mass of this galaxy.
In the following, therefore, we physically associate this metal-line
system with the bright, early-type spiral first identified by 
\cite{spinrad93}.  Figure~\ref{fig:1671} shows our WFCCD spectrum 
of this galaxy.  One identifies strong Ca\,H+K, G-band, and Balmer
absorption lines in the spectrum and also a weak but significant 
[OII] emission line.  The spectral type is that of an Sb galaxy
\citep{kennicutt92} and the Balmer lines suggest a significant
episode of star formation $\approx 1$\,Gyr ago.

\begin{figure*}[ht]
\begin{center}
\includegraphics[angle=90,width=6.5in]{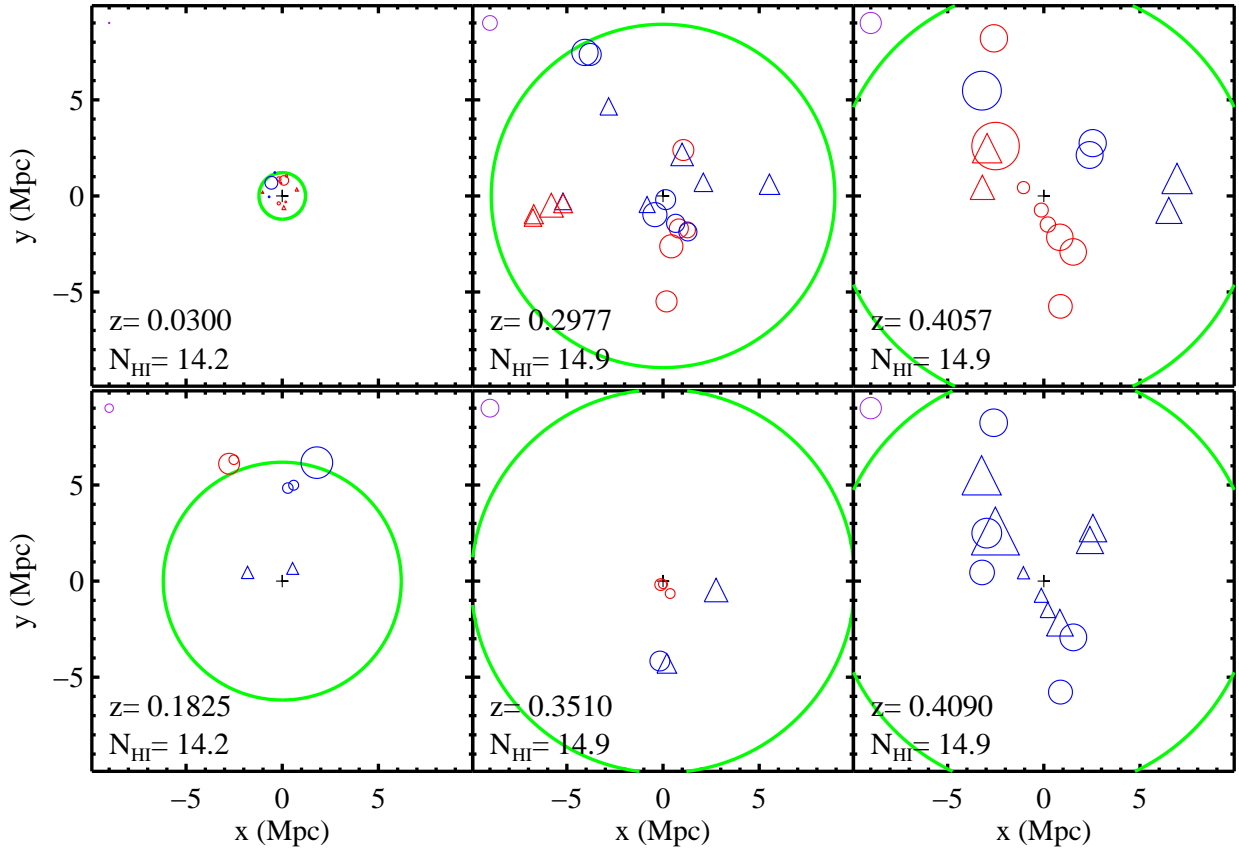}
\caption{Plot of the spatial distribution of \lya\ `clouds'
with $\mnhi > 10^{14} \cm{-2}$ but no apparent
metal-line absorption.
The symbols and color coding are the same as Figure~\ref{fig:galx}.
}
\label{fig:nmgalx}
\end{center}
\end{figure*}

This identification has several striking implications.
First, the observations suggest a tenuous, ionized multi-phase gas
within the halo of the early-type spiral.
As discussed in \cite{chen00}, the absorption system exhibits
two distinct phases: 
(1) a photoionized gas associated with Si$^+$, Si$^{++}$, C$^+$ ions
and presumably the majority of \ion{H}{1} gas.  The temperature
of this phase is a few 10$^4$\,K.
Adopting the mean extragalactic background
radiation field inferred by \cite{dave01}, the mean number density of the
gas is $n_H = 6 \sci{-4} \cm{-3}$; 
(2) a hot, ionized gas associated with \ion{O}{6}, \ion{N}{5},
and probable \ion{S}{6} absorption (Paper~I).
It is likely this gas is collisionally ionized with temperature
$T \approx 2\sci{5}$\,K.  
It is important to note that the photoionized gas may have 
characteristics similar to higher redshift \ion{Mg}{2} absorbers.
The association of this gas with the bright, early-type spiral is 
suggestive of the observed connection between bright spirals
and \ion{Mg}{2} absorbers at $z \sim 1$. 
Indeed, \cite{spinrad93} reported a tentative detection of the
\ion{Mg}{2} doublet and noted the resemblance between this absorber
and \ion{Mg}{2} systems \citep{steidel93}.
Adopting the photoionization model from Paper~I, we predict
$\log \N{C\,IV} > 13.5$ from the low-ionization phase alone.
We expect a \ion{C}{4} equivalent width in excess of 0.3\AA, 
and, therefore, that this absorber would follow the results of
other \ion{C}{4}/galaxy absorber-pairs \citep{chen01}.

Second, the velocity offset between the absorber and galaxy redshift is
small ($\delta v \approx -15 \mkms$) suggesting the gas is 
bound to the galaxy.
Third, it suggests a significant mass of gas in the outskirts
of the galactic halo.  Applying the ionization corrections
derived in Paper~I, we derive a total hydrogen column density
$N_H = 10^{19} \cm{-2}$ for the gas associated with the lower 
temperature phase.  
If we assume a constant surface density to $\rho = 100$\,kpc
with covering fraction $f$, then the integrated mass is 
$M = f \cdot 3 \times 10^9  \msol$.  
If the surface density increases radially toward the galactic
center, then the gas mass could be considerably higher even if 
the medium is clumped.  
Fourth, the gas has nearly solar metallicity: [Si/H]~$=-0.25$\,dex.  
For this reason, it is very unlikely the gas is being accreted
from the local IGM.   Instead,
we contend the gas was enriched within the galaxy
and then expelled to large radii.
This interpretation is supported by the strong Balmer absorption
lines in the galaxy spectrum (Figure~\ref{fig:1671}) which indicate
a recent episode of star formation.
Assuming a distance of 100\,kpc, the gas would only need an outflow
velocity of $\sim 100 \mkms$ to travel this distance in 1\,Gyr.

Fifth, we emphasize that the \ion{O}{6} equivalent width is 
among the largest
observed at extragalactic distances.  
In \cite{chen00}, we speculated that this gas 
may arise in an intragroup medium.  The results of our galaxy survey 
contradict this claim.  Instead, one must associate the gas with 
one of the two nearby galaxies.  This has further implications for the
multi-phase nature of the absorber.  If the two phases are co-spatial, then 
it is very likely in pressure equilibrium with the photoionized gas
embedded within the hotter, collisionally ionized \ion{O}{6} gas.
Adopting $n_H = 6\sci{-4} \cm{-3}$ and $T \approx 20000$\,K for the
photoionized phase and also $T_{hot} = 2\sci{5}$\,K for the
\ion{O}{6} phase, we derive $n_H^{hot} = 6 \sci{-5} \cm{-3}$.  As
\cite{chen00} noted, the gas must have a density on this order or lower
to avoid cooling in less than a dynamical time.
Therefore, a qualitatively self-consistent model can be constructed
where this
partial Lyman limit system is a pressure-supported photoionized `cloud'
located in a warm, collisionally ionized gas at large radii from this
early-type spiral.

\begin{table*}[ht]\footnotesize
\begin{center}
\caption{{\sc GALACTIC ENVIRONMENT SUMMARY\label{tab:galsumm}}}
\begin{tabular}{cccccccccl}
\tableline
\tableline
$z$ & $N_{\rm HI}$ & $N_{\rm OVI}$ & [M/H] & 
$\rho_{nearest}^a$ & $L_{nearest}^a$ & 
$L_{lim}^b$ & $\rho_{>L_*}$& $n_{\rm 3Mpc}^c$ & Comment \\
 & & & & (kpc) & ($L_*$) & ($L_*$) & (kpc) & &   \\
\tableline
%\cutinhead{Metal line Systems}
0.0918&$14.5$&$>-1.4$&$13.8$& 412&  0.1&  0.1& 2354&   7(1.2)& Filament?\\
0.0965&$14.7$&$-1.5$&$13.7$& 247&  0.3&  0.1&  351&   5(1.2)& Group\\
0.1671&$16.5$&$-0.2$&$14.8$& 108&  4.1&  0.4&  108&   1(1.8)& Galaxy\\
0.1829&$14.1$&$>-2.0$&$14.0$&4859&  0.7&  0.5& 6433&   0(1.9)& Void?\\
0.3608&$15.1$&$>-0.7$&$<13.3$& 220&  8.2&  2.2&  220&   5(0.4)& Group\\
0.4951&$14.4$&$>-0.3$&$14.3$&9378&  5.8&  4.7& 9378&   0(0.0)& ??\\
\tableline
%\cutinhead{Non-Metal line Systems}
0.0300&$14.4$&$<-1.5$&$<13.5$& 360&  0.0&  0.0&  888&   2(0.8)& Group?\\
0.1825&$14.9$&$<-1.5$&$<13.8$&4851&  0.7&  0.5& 6421&   1(1.9)& Void\\
0.2977&$14.0$&$<-1.5$&$<13.3$& 236&  2.6&  1.4&  236&  10(0.3)& Halo\\
0.3510&$14.2$&$<-1.5$&$<13.5$& 218&  0.9&  2.1& 4163&   1(0.4)& Void?\\
0.4057&$14.9$&$<-2.0$&$<13.6$& 747&  1.3&  2.9&  747&   3(0.0)& Filament?\\
0.4089&$14.4$&$<-1.5$&$<13.5$&3242&  4.0&  3.0& 3242&   3(0.0)& Filament?\\
\tableline
\end{tabular}
\end{center}
\tablenotetext{a}{Impact parameter and luminosity of the 
nearest galaxy to the sightline.}
\tablenotetext{b}{Limitting magnitude assuming $R_{lim}$=20mag.}
\tablenotetext{c}{Number of galaxies with physical impact parameter
 $<3$\,Mpc, $L > $ max$(L_*/2,L_{lim})$, and $|\delta v| < $750\kms.  The quantity in parenthesis is
 the average number using the volume of this 3\,Mpc radius region
 and the luminosity limit and also adopting the survey completeness to an angular diameter distance 3\,Mpc given
 by Figure~\ref{fig:complete} which is dominated by the non-overlapping nature of the flanking fields.}
\end{table*}

Sixth, adopting the $n_H^{hot}$ density and assuming the hot
phase fills the galactic halo to $r = 100$\,kpc, we derive
a hot gas mass $M^{hot} \approx 5 \sci{9} \msol$.  
Seventh, we note that this system has the highest \nhi\ value
along the PKS0405--123 sightline and is the 
only absorber unambiguously identified with a galactic halo.
While \lya\ `clouds' with $\mnhi > 10^{14} \cm{-2}$ are strongly
correlated with galaxies (Paper~II), the gas may be predominantly
located in the local environment of these galaxies as opposed to
individual halos.

Finally, the identification of this \ion{O}{6} absorber 
with a single galactic
halo is an important result in light of cosmological models
of \ion{O}{6} gas.  We emphasize that this \ion{O}{6} absorber has one of
the largest equivalent widths at extragalactic distance.  Cosmological
simulations suggest that absorbers with large equivalent widths are the most
likely to be collisionally ionized and also related to large-scale
structures \citep{dave01b}.  In this case, however, the gas 
appears related to an individual galactic halo and therefore bears
greater resemblance to the \ion{O}{6} gas observed in the halo 
of the Milky Way \citep{sembach03}.  
It will be very valuable to examine the galactic environment of
the strongest \ion{O}{6} absorbers and perform quantitative
comparisons with simulations.

\subsection{$\mnhi > 10^{14} \cm{-2}$ Absorbers Without Detected Metal-Lines}
	
Before discussing the trends in the galactic environments of the
metal-line absorbers, consider the environment of the absorbers
with comparable \nhi\ but without detected metal-lines.
Figure~\ref{fig:nmgalx} presents a summary of the galactic environment
of the non-metal systems with $\mnhi > 10^{14} \cm{-2}$ from 
\cite{williger05}.
Examining the Figure, we note a picture qualitatively similar to the
results for the metal-line systems.  There are examples of dense
galactic environments ($z=0.0300, 0.2977, 0.4057$), absorbers potentially
within individual galactic halos ($z=0.2977,0.3510$), 
and absorbers in very sparse environments ($z=0.1825$) 
These results indicate that the cross-correlation observed
between \lya\ forest clouds with $\mnhi > 10^{14} \cm{-2}$
and galaxies is not dominated by the metal-line systems.

\subsection{Emerging Trends?}

Considering the metal-line systems as a single class of absorbers,
one might have expected similarities in their galactic environments.
As Figure~\ref{fig:galx} qualitatively
shows (summarized in Table~\ref{tab:galsumm}), 
the absorption systems exist near or within a wide range of 
galactic environments.
There are examples where one can associate the absorber with the
halo of a nearby bright galaxy, examples where there is no
bright host yet a group or large-scale structure, and 
even an example ($z=0.1825$) with no associated galaxies with
$L \gtrsim L_*/2$.  It is evident that metal-enriched gas
generally associated with the IGM
is located in a wide variety of environments at low redshift.
%One implication is that the enrichment processes of the IGM
%occur in nearly all galactic environments.
A focus of our future work based on over $10$ quasar fields
will be to examine the incidence of \ion{O}{6} absorbers in
each of these environments.

Although the PKS0405$-$123 sightline contains only a small sample
of metal-line absorbers, one can search for emerging trends between
their galactic environment and absorber properties. 
As a first step, we might separate the absorbers according to
whether they are located in `rich' or `poor' galactic environments.
To qualitatively assess the environment, we count the number of
galaxies with $L>L_*/2$ and $|\delv| < 750\mkms$ within 3\,Mpc 
of the sightline (Table~\ref{tab:galsumm}).
Under this definition, we classify the
$z=0.0918, 0.0965, 0.36$ absorbers as `rich' ($N_{gal} > 4$)
and the $z=0.167, 0.1825$ absorbers as `poor' ($N_{gal} < 3$).
The survey is not sufficiently deep to consider the system at $z=0.4951$.
With these classifications, there is no obvious trend between
galactic environment and absorber properties.  The $z=0.167$ and
$z=0.1825$ absorbers have very different characteristics
(\nhi\ value, metallicity and ionization states).  The
rich-environment 
$z=0.0965$ and $z=0.3608$ absorbers also have dissimilar 
physical properties.
In short, we find no trend between galactic environment and
metallicity, \nhi, or ionization state, contrary to the strong
correlation between $N({\rm H\,I})$ and galaxy number density 
reported by Bowen et al.\ (2002).
Given the small sample size (especially metal systems at $z < 0.1$
where $L < L_*/10$), our conclusions do not necessarily
disagree with the findings of \cite{stocke06} who argue that the
presence of
metal-line systems is significantly correlated with galactic
environment, but do not draw a conclusion about the properties
of those systems with environment.

Examining Figure~\ref{fig:galx}, it is apparent that
the metal-line systems, which show a two magnitude spread in H\,I column
density and metallicity, arise in regions with diverse galaxy
density and distribution.  For example, contrast the absorbers at $z=0.0918$
and $z=0.0965$ which exhibit similar H\,I column density and metal-line 
absorption.  In both cases, there is a significant excess of galaxies
observed at the redshift.  In the case of $z=0.0965$, the majority of
galaxies lie within 1~Mpc of the absorber and very likely form a virialized
group. In contrast, the galaxies associated
with the $z=0.0918$ absorber cover the entire field-of-view in a distribution
suggesting a larger-scale, unvirialized structure (e.g.\ filament or sheet).

\subsection{The $z\sim 0.05$ \lya\ Void}

\cite{williger05} have stressed that the sightline to PKS0405--123 
exhibits no \lya\ absorbers with $\mnhi > 10^{14} \cm{-2}$ between
$0.0320 < z < 0.0814$.  They also have
argued that the probability of this
occurrence from a random redshift distribution is $P<0.0004$.
Given the correlation between strong absorbers and galaxies
\citep{chen05}, one might expect a similar void in galaxies in
this redshift interval.
In the inner $10'$ minute field, we identify four galaxies 
with redshift $0.0320 < z < 0.0814$ at redshift:
$z_{gal}=$.043 (ID~2716), 
0.0668 (ID~2999), 0.0767 (ID~1463), 0.0770 (ID~993).
Adopting the SDSS $r^*$ luminosity function \citep{blanton05},
we predict $7.4 \pm 3.2$ galaxies in the volume subtended by
a $10'$ radius where the uncertainty includes the effects
of galaxy clustering
\citep{connolly02}.  Therefore, the number of observed galaxies is below the
expected central value but well within the expected
scatter.  
%It is notable, however,
%that the galaxies all have impact parameter $\rho > 300$\,kpc,
%magnitudes $R > 18.9$\,mag (i.e.\ $L < L_*/2$), and late-type
%spectral features.

\section{Summary}

We summarize the principal results of this paper as follows:

\begin{itemize}

\item  We presented the results of a galaxy redshift survey performed with
the WFCCD spectrometer of the field surrounding PKS0405--123.
The survey is nearly complete to $R=20$\,mag within a radius
of 10$'$.  The Survey is supplemented by flanking fields covering
$\approx 1 \sqdeg$ to $R=19.5$\,mag.  All of the
galaxy spectra are publically available at 
http://www.ucolick.org/$\sim$xavier/WFCCDOVI/.

\item We examined the galactic environment of the metal-line
absorption systems identified along the sightline to PKS0405--123,
all but one of which show \ion{O}{6} absorption.
The systems arise within or near regions of the universe with
a wide range of galactic environments including individual galactic
halos, galaxy groups, filamentary-like structures, and regions 
devoid of bright galaxies.  This suggests that metal-enrichment
occurs in galaxies with a wide range of luminosity and environment.

\item For the sightline to PKS0405--123, there is no significant difference
in the galactic environment of $\mnhi > 10^{14} \cm{-2}$ absorbers with 
or without metal-lines.  If this holds with larger samples, it
would suggest that metal-enrichment is a stochastic process with
little dependence on galactic environment.

\item In future papers, we will increase the sample size of
sightlines by nearly an order of magnitude.  Particular emphasis
will be placed on studying the galactic environment of low
redshift, \ion{O}{6} absorbers.

\end{itemize}

\acknowledgments

We thank D. Kelson and S. Burles
who inspired several aspects of the WFCCD pipeline.
We thank M. Geha for providing observations of one mask
for PKS0405--123.
This work was partially supported by a FUSE GI grant to
JXP under NASA contract NAG5-12496.

%\clearpage 

\clearpage

\clearpage

\end{document}